\documentclass[twocolumn,aps,pra]{revtex4}
\usepackage{color}
\usepackage{graphicx}
\usepackage{bm}
\usepackage{amssymb}

\def\bea{\begin{eqnarray}}
\def\eea{\end{eqnarray}}
\def\be{\begin{equation}}
\def\ee{\end{equation}}

\begin{document}

\title{Buffer gas induced collision shift for the $^{88}$Sr $\bf{^1S_0-^3P_1}$ clock transition}

\author{Nobuyasu Shiga$^1$}
\author{Ying Li$^1$}
\author{Hiroyuki Ito$^1$}
\author{Shigeo Nagano$^1$}
\author{Tetsuya Ido${}^{1,2}$}
\email[Corresponding author: ]{ido@nict.go.jp}

\affiliation{$^1$Space-Time Standards Group\, National Institute of Information and Communications Technology \\
$^2$CREST, Japan Science and Technology Agency \\
4-2-1 Nukui-kitamachi\, Koganei\, Tokyo,\ 184-8795, Japan
}

\author{Katarzyna Bielska}
\author{Ryszard S. Trawi\'nski}
\author{Roman Ciury{\l}o}
\affiliation{Instytut Fizyki, Uniwersytet Miko\l aja Kopernika, ul. Grudzi\c{a}dzka
5/7, 87-100 Toru\'{n}, Poland}

\begin{abstract}     
Precision saturation spectroscopy of the $^{88}{\rm Sr}\ ^1S_0-^3P_1$ is performed in a vapor cell filled with various
rare gas including He, Ne, Ar, and Xe. By continuously calibrating the absolute frequency of the probe laser,
buffer gas induced collision shifts of $\sim $kHz are detected with gas pressure of 1-20 mTorr.
Helium gave the largest fractional shift of $1.6 \times 10^{-9} {\rm\ Torr}^{-1}$. Comparing with a simple impact
calculation and a Doppler-limited experiment of Holtgrave and Wolf [Phys. Rev. A {\bf 72}, 012711 (2005)],
our results show larger broadening and smaller shifting coefficient, indicating effective atomic loss due
to velocity changing collisions. The applicability of the result to the $^1S_0-^3P_0$ optical lattice clock
transition is also discussed.
\end{abstract}
\date{\today}

\maketitle

\newcommand{\LevelExpSetup}[1][\w]{
\begin{figure}
\includegraphics*[width=3.5in]{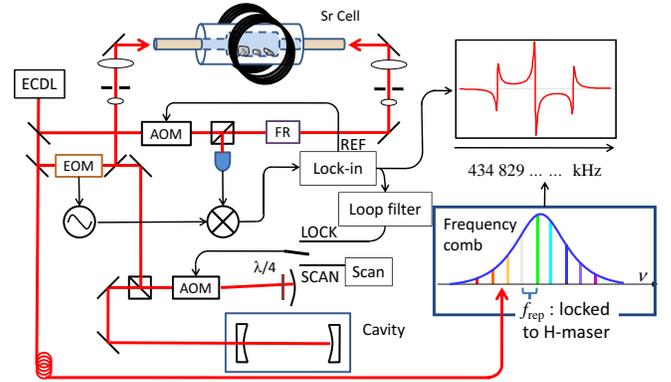}\hfill
\caption{(Color online) Experimental configuration of the precision saturated absorption spectroscopy.
ECDL: extended cavity diode laser, FR: Faraday rotator, EOM: Electro-optic modulator,
AOM: acousto-optic modulator}
\label{Fig:ExpSetup}
\end{figure}
}

\newcommand{\DispersionAllan}[1][\w]{
\begin{figure}
\includegraphics*[width=3.5in]{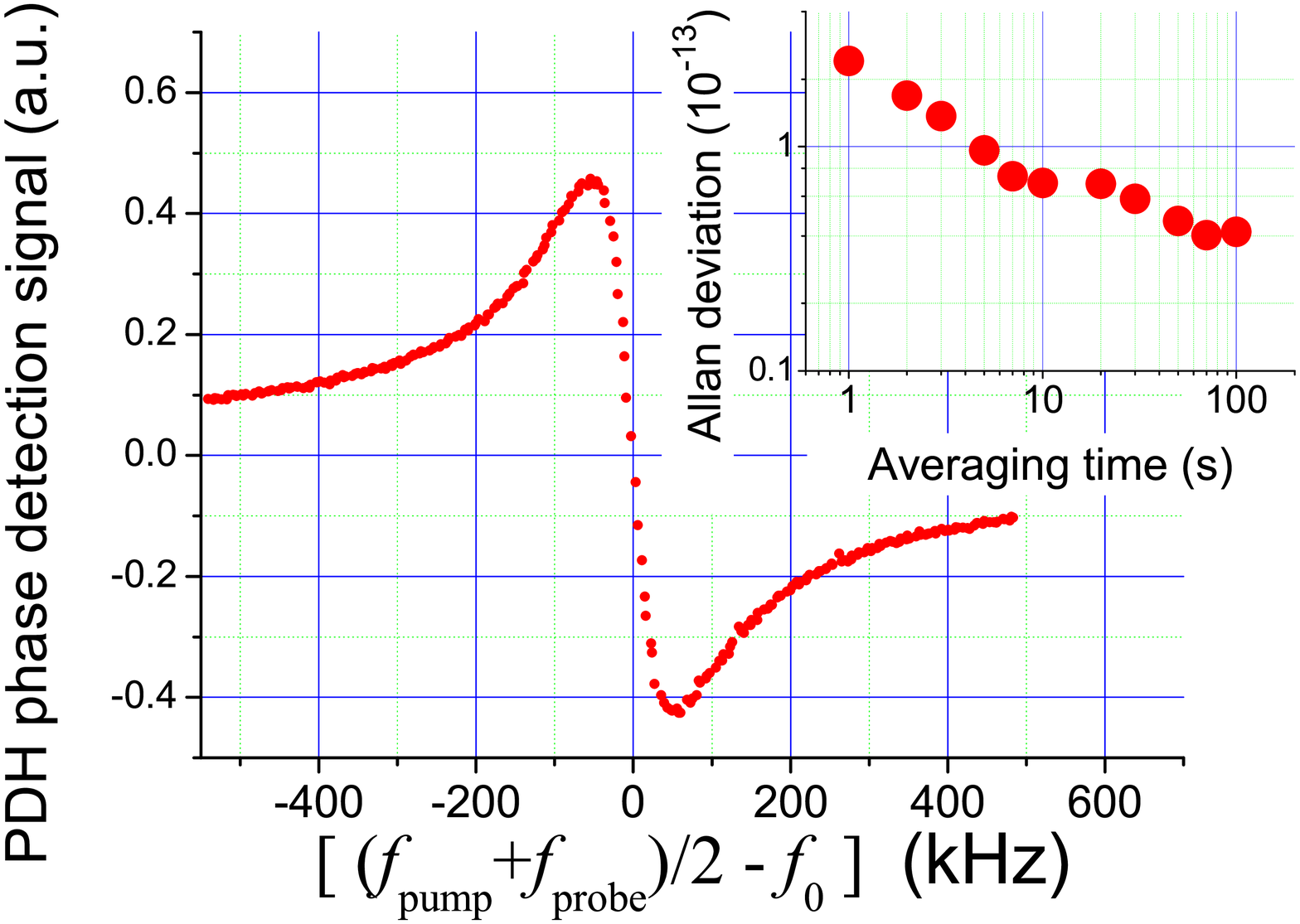}\hfill
\caption{(a)Dispersion signal of FM side band spectroscopy. The offset frequency of $f_{0}=$434 829 121 312 kHz
is the true atomic resonance measured in \cite{idoPRL}.
(b) Modified Allan standard deviation of the 689 nm laser frequency tightly locked to the resonance signal obtained
by the saturation spectroscopy.}
\label{Fig:DispCurve}
\end{figure}
}

\newcommand{\HighlightGraph}[1][\w]{
\begin{figure}
\includegraphics*[width=3.75in]{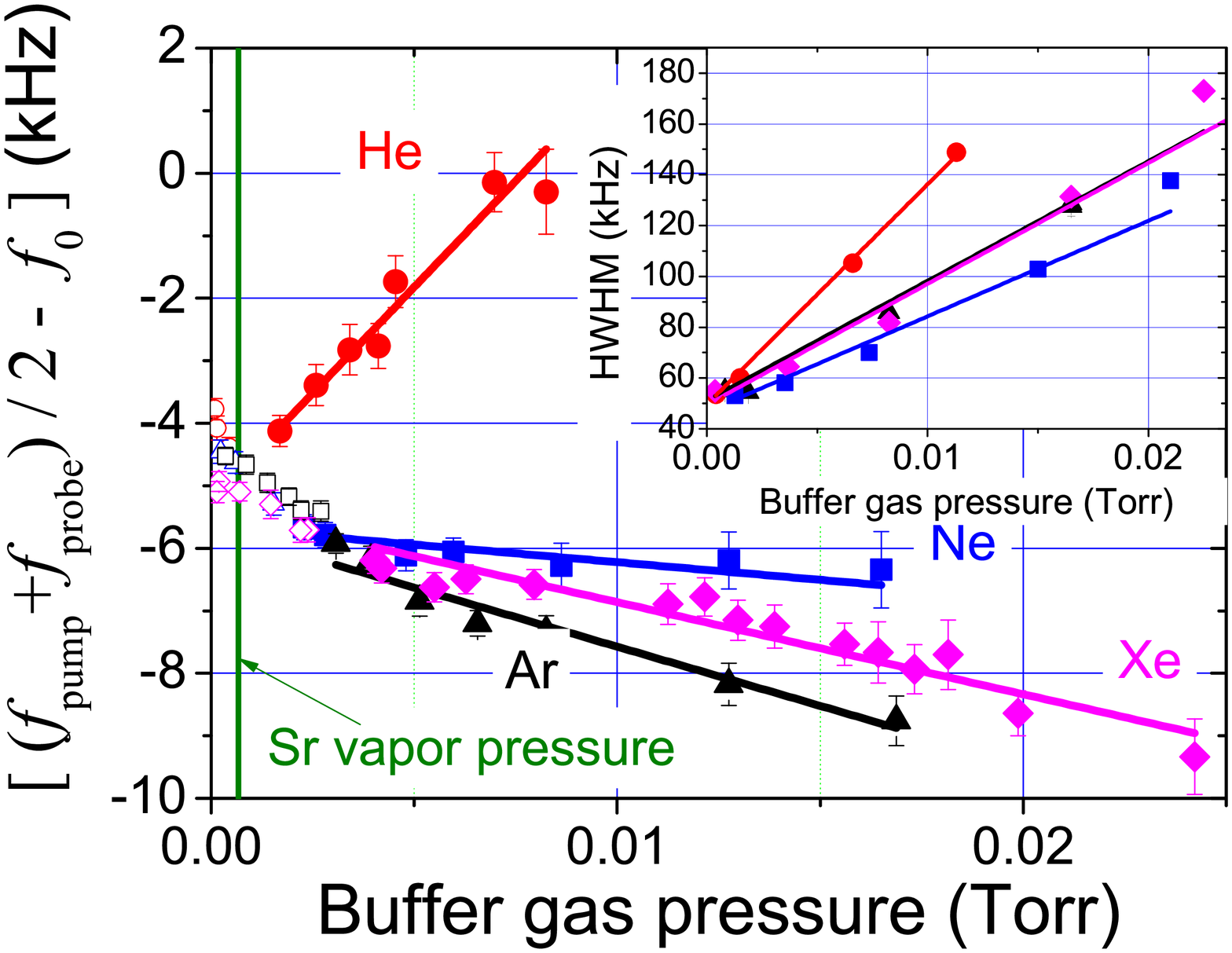}\hfill
\caption{(Color online) Shifts of the resonance frequency induced by various kinds of buffer gas.
The corner of the slope around 3 mTorr is attributed to the inhomogeneous pressure distribution in the cell,
which occurs due to insufficient buffer gas pressure relative to strontium vapor. Inset is the broadening
of the Doppler-free atomic resonance.}
\label{Fig:GammaShift}
\end{figure}
}

The latest progress of optical clocks is so rapid that their capability is surpassing that of microwave clocks. A fractional accuracy of $\sim 10^{-17}$  was demonstrated with single-ion systems \cite{TilScience}.
This remarkable precision, on the other hand, requires us to consider various
 systematic shifts which have been so far ignored. Neutral atomic clocks, particularly fountain
clocks, have been often limited in their accuracy by binary collisions \cite{Rbfountain}. Besides that, whether the clock is
neutral-atom or ionic system, collision shifts induced by background gas (BG) \cite{Allard82} may ultimately limit the accuracy in this unprecedented level.
However, the characterization of BG induced collision shifts is difficult in real clocks because the deliberate injection of buffer gas for the shifting of  atomic resonance normally causes loss of atoms, limiting the coherence time needed to evaluate subtle BG induced shifts.

From the beginning of laser spectroscopy, frequency shifts and broadening caused by background gas collisions
have been investigated both experimentally and theoretically \cite{Allard82}. The spin-forbidden transitions
$^1S_0-^3P_1$ of atoms with two outer electrons have been major testing grounds due to their fairly narrow natural
linewidths and level simplicity free from hyperfine structure.
Experiments were done for Ca \cite{Crane94}, Sr \cite{Crane94,Tino92,Holtgrave05},
Cd \cite{Dietz80,Bielski02}, Hg \cite{Jacobs03,Anderson07} and Yb \cite{Kimball99} with perturber pressures
of $10^0-10^2 {\rm\ Torr}$. In these experiments, shifts and broadening exceeds 10 MHz, and high order effects
such as collision time asymmetry or speed-dependence of collisional broadening and
shifting can be observed \cite{Shannon86,Bielski00}. The higher-order effects, however, disturb the accurate
measurement of the linear collision shift and broadening, which is necessary
to evaluate systematic shifts of atomic clocks.

In this paper, buffer gas induced collision shifts for the $^{88}{\rm Sr}\ ^1S_0-^3P_1$ transition is reported based on precision saturation absorption spectroscopy with the FM sideband technique \cite{I2Jan,Bjorklund83}. Continuously monitoring the probe laser
frequency using a frequency comb referenced to International Atomic Time (TAI), we compared the resonance
frequency with the true atomic resonance measured with ultra-cold atoms \cite{idoPRL}.
The detected background gas induced collision shift was $\sim {\rm kHz}$, which is, to the best of our knowledge,
the finest definitive measurement of optical frequency shifts induced by foreign-gas collisions.

\LevelExpSetup

The $(5s^2) ^1S_0 - (5s5p)^3P_1$ transition at $\lambda =689\;{\rm nm}$ has a natural line width
$\gamma = 2 \pi \times 7.4\;{\rm kHz}$ 
 which corresponds to the saturation intensity
$I_s = 2\pi ^2 \hbar c \gamma / 3\lambda^3 = 2.9\;\mu\rm{W/cm}^2$, where $c$ and $2\pi \hbar$ is the
speed of light and the Planck constant, respectively. The experimental setup of our saturated absorption
spectroscopy is shown in Fig. 1.
The spectral width of an extended cavity diode laser (ECDL) is reduced to less than 500 Hz
by locking the laser frequency to a cavity with a basic Pound-Drever-Hall (PDH)
method \cite{PDH}. The modulation index and frequency $\omega_M$ are 0.17 and $2\pi \times 21.7 {\rm MHz}$, respectively.
This modulation is used for the saturation absorption spectroscopy as well. Strontium is contained
in a heat pipe with Brewster window plates at both ends. The typical temperature of the pipe is
$728\;{\rm K}$.
The two ends of the active area (20 cm length) are water-cooled so that the strontium is not deposited on the windows.
The buffer gas is introduced from one end of the pipe, and Pirani gauges are installed at both ends for the confirmation
 of uniform pressures in the tube. The magnetic sublevels are well defined by a bias magnetic field of 3 G produced by a Helmholtz coil pair. Pump and probe beam of
the saturation spectroscopy have the same linear polarization which excites the $\pi$ transition of
$\Delta m_J=0$. Pinholes are placed in both arms to overlap the pump and probe beam. The peak intensity and $1/e^2$
radius of the pump
beam (probe beam) are $270\;\mu{\rm W/cm}^2$ and 4.2 mm ($50\;\mu{\rm W/cm}^2$ and 3.3 mm), respectively.
After interacting with Sr vapor, the polarization of the probe beam is rotated 90 degrees and coupled out
by a polarization beam splitter. The pump beam is frequency shifted by 80 MHz from the probe
beam in order to avoid DC noise caused by the interference between the two beams. In addition, the pump beam
is chopped with a frequency of 5 kHz to extract the
saturation signal induced by the pump beam.
The signal is first demodulated at $\omega _M$ and then lock-in detected. The absolute frequency of the laser is always
monitored by a frequency comb referenced to International Atomic Time (TAI).

A sample of PDH phase detection signal with helium buffer gas of $4\times 10^{-4}{\rm\ Torr}$ is shown in Fig. \ref{Fig:DispCurve}(a).
The absolute frequency of the probe laser is expressed as a differential frequency $(f_{\rm pump}+f_{\rm probe})/2 - f_0$,
where $f_{\rm pump}$ and $f_{\rm probe}$ is pump and probe laser frequency, respectively, and the true atomic resonance $f_0$ is the frequency precisely measured in \cite{idoPRL} from ballistically
expanding ultracold atoms. The full width half maximum (FWHM) is nonlinear-fitted to be 107 kHz, which agrees
with saturation broadening (80 kHz) and transit time broadening (31 kHz) expected from the beam intensity and
diameter. This dispersion shape works as an error signal to lock the laser.
The stability of this Sr vapor cell clock is shown in Fig.~\ref{Fig:DispCurve}(b) as modified Allan deviation
against a Hydrogen maser. The bottom of the instability is $4 \times 10^{-14}$ at 100 s.
The stability is limited by fluctuations of buffer gas and vapor pressure or pump beam
intensity that changes the light-pressure-induced lineshape asymmetry \cite{Grimm}.

\DispersionAllan

We measured the pressure dependence of the background gas induced broadening and shift for various
background gases. The buffer gases are rare gases, namely helium, neon, argon, and xenon. The attempts to fill with nitrogen failed since the gas was quickly adsorbed by solid strontium. The broadening is determined
from the dispersion curve as in Fig. \ref{Fig:DispCurve}(a). Locking the laser frequency to the zero-crossing of dispersion curve, the resonance is determined by a frequency
comb that is referenced to TAI.  We determined the slope of the collisional broadening and the frequency shift as
a function of the buffer gas pressure by doing a least square fit, and the result is summarized
in Table \ref{Tab:ExpTheo}.

The frequency shift data are summarized in Fig. \ref{Fig:GammaShift}. The experimental condition described above has
resulted in  a residual offset of -4 kHz from the true atomic resonance. Part of the reason is a shift caused by
homonuclear two body collisions. By changing the temperature of the heat pipe, we measured the Sr-Sr collision
shift coefficient to be $-1.3 \times 10^{-10} {\rm\ Hz\ cm^3}$ \cite{densityshift}. This effect causes a shift of -1 kHz
for the condition of Fig. 3. We suspect that the most probable reason for the remaining -3 kHz is residual Doppler shifts.
The spatial filters in the two arms still permit a mutual beam tilt of $\theta = 62\ \mu {\rm rad}$. In the case that atoms
are directed as in an atomic beam, this tilt causes Doppler shift of
$k_L v_{\rm m} \sin \left( \theta /2 \right) = 15{\rm\ kHz}$ maximum, where $k_L$ is the wave number of the light,
$v_m=\sqrt{2k_B T/m} = 371 {\rm\ m/s}$ is most probable speed of Sr atoms
at gas temperature $T=455^\circ {\rm C}$, $m$ is the Sr mass and $k_{B}$ is the Boltzmann constant.
Our spectroscopy is ideally performed with an isotropic velocity distribution of atoms. However, a slightly biased velocity
distribution may be produced in the heat pipe due to an inhomogeneous temperature distribution of the pipe.
Since the beam passes above bulk solid strontium, the distribution could be biased upward.
Other possibilities are light-pressure-induced lineshape asymmetry \cite{Grimm} and recoils due to photon absorption and emission
\cite{HgSYRTE, TaraCa, ChrisCa}.

Only the collisions with He show a positive shift coefficient, whereas Ne, Ar,
Xe show negative coefficients. The open symbols at lower pressures are excluded from the fitting since the vapor pressure of strontium is not small enough comparing with the buffer gas pressure.
Even slightly above the vapor pressure, the buffer gas pressure wasn't uniform at pressures below $3 \times 10^{-3} {\rm\ Torr}$.
We added the buffer gas from one side of the pipe, and the spatial gradient of the gas pressure was observed by Pirani gauges placed at the both ends of the heat pipe.

\HighlightGraph

The collisional width $\Gamma$ (HWHM) and shift $\Delta$ in the impact
approximation are proportional to the density of perturbers $N$ and are given by the well-known expression
\cite{Baranger58,Allard82}:
\begin{equation}
\label{Eq:WidthShift}
\Gamma+i\Delta=N\int d^{3}\vec{v}_{r} f_{\mu}(\vec{v}_r) v_{r} \int_{0}^{\infty} d\rho\; \rho
\left\{1-\left< S_{ii} S_{ff}^{-1}\right>_{Av.}\right\}
\end{equation}
where $f_{\mu}(\vec{v}_r)$ is the Maxwellian distribution of relative velocities $\vec{v}_r$
of colliding absorber and perturber having reduced mass $\mu$, and $\rho$ is the impact parameter.
In this work we report results obtained in the mean speed approximation in which
calculations are done just for the mean relative speed $\bar{v}=\sqrt{8 k_{B} T/(\pi\mu)}$ of colliding atoms.
For the intercombination transition $^{1}S_{0}-^{3}P_{1}$,
in the adiabatic approximation the angular average of scattering
matrix elements in the initial and final states
$\left< S_{ii}S_{ff}^{-1}\right>_{Av.}=\frac{1}{3}e^{-i\eta_{A-X}}+\frac{2}{3}e^{-i\eta_{B-X}}$
can be expressed in terms of phase shifts $\eta_{A-X}$ and $\eta_{B-X}$ induced by the quasimolecular
transitions $A^{3}0^{+}-X^{1}0^{+}$ and $B^{3}1-X^{1}0^{+}$, respectively \cite{Bielski00,Bielski04}.
These phase shifts are integrals
$\eta=\int_{-\infty}^{+\infty}dt \Delta V\left(r(t)\right)/\hbar$
of the excited and ground electronic state potentials differences
$V(A^{3}0^{+})-V(X^{1}0^{+})$ and $V(B^{3}1)-V(X^{1}0^{+})$
for Sr--Rg(He, Ne, Ar, Xe) systems.
For simplicity we assumed straight
line trajectories with the interatomic separation $r(t)=\sqrt{\rho^{2}+v_{r}^{2}t^{2}}$.

The interaction potentials are approximated here by a modified Lennard-Jones formula
$V(r)=C_{12}/r^{12}+C_{8}/r^{8}+C_{6}/r^{6}$.
Following Hindmarsh et al. \cite{Hindmarsh67} $C_{12}$ coefficients are estimated using Hindmarsh radii
$r_{H}({\rm Rg})$ for rare gases Rg \cite{Bielski80} and $r_{H}({\rm Sr})$ for Sr.
The Hindmarsh radii $r_{H}({\rm Sr})$ have been estimated using the charge densities calculated by Mitroy and Zhang
\cite{Mitroy09}. They were found to be $9.4773\; a_{0}$ and $10.8801\; a_{0}$  in the ground $^{1}S_{0}$ and
excited $^{3}P_{0,1,2}$ electronic state of Sr atom, respectively.
The atomic unit of length is $a_{0}\approx 0.05292\;{\rm nm}$.
The dispersion coefficients $C_{6}$ and $C_{8}$ for Sr--Rg were also calculated by Mitroy and Zhang \cite{Mitroy09}.
The dispersion coefficient given for excited states
$^{3}\Sigma^{+}$ and $^{3}\Pi$ were converted to appropriate coefficients for
states $A^{3}0^{+}$ and $B^{3}1$ using the following asymptotic relations:
$C_{6}(A^{3}0^{+})=C_{6}(^{3}\Pi)$ and $C_{6}(B^{3}1)=\frac{1}{2}C_{6}(^{3}\Sigma^{+})+\frac{1}{2}C_{6}(^{3}\Pi)$.
The term $C_{8}/r^{8}$ included in our calculations allows for a better description of
the interaction anisotropy in the excited state of Sr.
We note that the addition of the $C_{8}/r^{8}$ and $C_{10}/r^{10}$ terms to the interaction
model changes the calculated broadening about a factor of two.
The change can be even bigger in the case of the shift and for Ne can lead to a change of its sign.
We also verified that if the thermal average is neglected,
the resulting error is less than 5\%.
The positive collisional shift observed for He  is caused by the dominating influnce of
the repulsive interaction $C_{12}/r^{12}$ on the collisional broadening and shifting.

\begin{table*}
\begin{center}
\caption{The comparison of our experimental pressure broadening and shifting coefficients with those
calculated in the impact approximation, as well as with experimental results obtained by
Holtgrave and Wolf \cite{Holtgrave05} in the Doppler limited regime and obtained by
Tino {\em et al.} \cite{Tino92} in the Doppler free regime.
All data are given in MHz/Torr and the one standard deviation uncertainties are given in parentheses.}
\label{Tab:ExpTheo}
\begin{tabular*}{13cm}{@{\extracolsep{\fill}} c|rrrr|rrrr}
\multicolumn{1}{c}{}&\multicolumn{4}{c}{Doppler limited}&\multicolumn{4}{c}{Doppler free}\\
\hline\hline
&\multicolumn{2}{c}{Experiment}&\multicolumn{2}{c|}{Impact calculation}&\multicolumn{2}{c}{Experiment}&\multicolumn{2}{c}{Impact calculation}\\
Perturber&\multicolumn{2}{c}{Holtgrave and Wolf}&\multicolumn{2}{c|}{($T=660\;{\rm K}$)}&\multicolumn{2}{c}{This work}&\multicolumn{2}{c}{($T=728\;{\rm K}$)}\\
&$\Gamma/p$&$\Delta/p$&\hspace{5mm}$\Gamma/p$&$\Delta/p$&$\Gamma/p$&$\Delta/p$&\hspace{5mm}$\Gamma/p$&$\Delta/p$\\
\hline
He&3.79(17)&+1.23(6)&3.89&+0.68&8.67(10)&+0.68(8)&3.68&+0.66\\
Ne&2.60(13)&-0.64(4)&1.17&-0.31&3.76(5)&-0.06(3)&1.10&-0.28\\
Ar&2.45(4)&-1.45(4)&3.30&-1.54&4.76(8)&-0.22(2)&3.05&-1.47\\
  &        &        &    &     &5.5(5)$^{a}$&        &    &     \\
Xe&3.44(23)&-1.51(3)&3.35&-1.36&4.71(6)&-0.12(1)&3.17&-1.33\\
\hline\hline
\end{tabular*}\\
\hspace{-8cm}$^{a}$Result obtained by Tino {\em et al.} \cite{Tino92}.
\end{center}
\end{table*}

Table \ref{Tab:ExpTheo} shows the comparison of the pressure broadening and shifting coefficients measured
in this work with those calculated in the impact approximation, Eq. (\ref{Eq:WidthShift}), and experimental
results obtained by other authors.
There is only one full set of data for rare gas pressure broadening and shifting of strontium
intercombination line reported by Holtgrave and Wolf \cite{Holtgrave05}.
Results of Ref. \cite{Holtgrave05} were obtained in the Doppler-limited regime at a temperature of
$660(47)\;{\rm K}$, where the magnitude of the broadening and shifts exceed 100 MHz.
Note that collisional broadening cross section determined by Holtgave and Wolf
agree very well with those obtained by Crane et al \cite{Crane94} for Ar and and differ by 23\% from
those for Ne. In Ref. \cite{Crane94} also the Doppler-limited spectroscopical method was used.
It is clear that our results do not agree with those obtained using Doppler limited spectroscopy
\cite{Holtgrave05,Crane94}. On the other hand our results agree well with those determined
by Tino {\em at al.} \cite{Tino92} for Ar in the Doppler free regime, see Table \ref{Tab:ExpTheo}.

The comparison of calculated results seems to give a bit better agreement with results obtained
in the Doppler-limited case. Clearly Doppler-free results are characterized by bigger collision broadening
coefficients and smaller magnitude of collision-shifting coefficients than those obtained from
the Doppler-limited case. This can be attributed to velocity changing collisions \cite{Berman79} which
change the absorber velocity significantly enough to push away the absorber from the condition
of Doppler-free resonance.
Therefore, such events of collisions which can cause large frequency shifts in the Doppler-limited regime
would not contribute to the signal in the Doppler-free case. Being a loss mechanism, it will lead to broadening
but not to shifting of the line.
This effect needs further theoretical and
experimental investigation. Moreover in these investigations it would
be very useful to carry out proper close-coupling broadening and shifting calculations \cite{Julienne86} based on
{\it ab initio} potentials.

The cold strontium system is currently studied extensively in many national standards institutes due to its application to
optical lattice clocks \cite{TakaNature}.
For the clock transition $^{1}S_{0}-^{3}P_{0}$, the interaction of the ground state rare gas atom
with the excited clock atom at $^{3}P_{0}$ state is described
by just one potential curve denoted $^{3}0^{-}$ approaching dissociation limit.
The dispersion coefficient describing this interaction fulfills the following
relation $C_{6}(^{3}0^{-})=C_{6}(^{3}\Pi)$
and with a good approximation is equal to $C_6(A^30^+)$ investigated here.
Moreover, in the case of collisions of clock atoms with background
gas atoms, the velocity-changing collisions will also play a crucial role leading to relatively large
collisional broadening and small collisional shifting. Therefore, results obtained here
as well as the discrepancy of the shift coefficients depending on Doppler-free or -limited system should help
to correctly estimate the impact of residual gas induced collision shift on optical atomic clocks.
The values of fractional shift coefficients $< 2\times 10^{-9} {\rm Torr^{-1}}$ obtained here can be treated as an upper limit for shifts caused
by the residual gas in optical lattice clocks which in fact should be even smaller.

We thank J. Mitroy and J. -Y. Zhang for their calculation of various $C_n$ numbers for strontium.
We acknowledge the fruitful comment and experimental help from W, Itano, M. Hosokawa, M. Kajita, A. Yamaguchi, M. Koide,
and H. Ishijima.
The theoretical part of this work was supported by the Polish MNISW (Project No. N N202 1489 33) as
a part of the program of the National Laboratory FAMO in Toru\'n, Poland.


\begin{thebibliography}{10}
\bibitem{densityshift}
This coefficient is 10 times less than that measured with cold atoms\cite{idoPRL}.
\end{thebibliography}
\end{document}